\documentclass[prl, aps, twocolumn, floats, showpacs, superscriptaddress]{revtex4-2}
\usepackage[makeroom]{cancel}
\usepackage{graphicx}
\usepackage{amsmath}
\usepackage{physics}
\usepackage{dsfont}
\usepackage{epstopdf}
\usepackage{amsfonts}
\usepackage{amssymb}
\usepackage{xcolor}
\usepackage{graphics} 
\usepackage{dcolumn}
\usepackage{bm}
\usepackage{array}
\usepackage{appendix}
\usepackage{nicefrac}
\usepackage{cellspace} 
\usepackage[left=2cm,right=2cm,top=2cm,bottom=2cm]{geometry}
\usepackage[normalem]{ulem}
\RequirePackage[
   hyperindex,colorlinks,bookmarksnumbered,
   plainpages=true,pdfstartview=FitH]{hyperref}
\hypersetup{linkcolor=blue,urlcolor=blue,citecolor=blue}

\begin{document}
\title{Time evolution in the multi-component dynamical Keldysh models}
\author{M. N.  Kiselev}
\email{mkiselev@ictp.it}
\affiliation{The Abdus Salam International Centre for Theoretical Physics,  Strada
Costiera 11, I-34151, Trieste,  Italy}

\begin{abstract}
We construct a time evolution operator for a dynamical quantum models with a multi-component random Gaussian field characterized by a long correlation time. We analyse three paradigmatic cases: non-Abelian $SU(2)$ symmetry class for the three-component Gaussian noise and Abelian symmetry classes $U(1)$ and $Z_2$ for the two- and one-component noises correspondingly.  Starting with a limiting case of the infinite correlation time (Keldysh model) we shown that statistics of the two-component random field with a non-zero mean is described by the Rice distribution. We derived and solved the system of coupled linear differential equations for the exact diagonal and off-diagonal  single-particle Green's functions. We showed how the finite time correlations in the Keldysh models are accounted by
a continuous fractions method.
\end{abstract}
\date{\today}
\maketitle
\noindent
{\it {\color{blue} Introduction.}} 
Original Keldysh model has been introduced by Keldysh \cite{keldysh1965,sadovskii2019}
to describe a 3-dimensional motion of an electron in a very smooth long-range Gaussian disorder potential with a variance $W^2$:
\begin{equation}
D({\bf r}-{\bf r'})\!=\!\langle V({\bf r})\!V({\bf r'})\rangle\! =\! W^2
\!\to\! D({\bf q})\!=\!(2\pi)^3W^2 \delta({\bf q}).\;\;\;
\end{equation}
It has been justified by a Borel re-summation of Feynman diagrams that the single particle retarded Green's function(GF)  $G^R$ of an electron with a dispersion $\epsilon({\bf p})$ can be obtained from a single particle GF {\it in a given realization} of the Gaussian disorder by averaging over all realizations:
\begin{equation}
G^R(E,{\bf p})=\frac{1}{\sqrt{2\pi W^2}}\int_{-\infty}^\infty\frac{d V \exp\left(-V^2/(2 W^2)\right)}{E-\epsilon({\bf p}) +V+i\delta}.
\end{equation}
This result has been re-derived by various diagrammatic methods, in particular, by solving exactly the Dyson equation in combination with the Ward identities \cite{efros1970, sadovskii2019,agd1963} and a path integral technique \cite{schmalian1998,
schmalian1999}.  The model got numerous applications in many branches of the modern condensed matter theory. The Keldysh model in its original formulation has been applied to the theory of heavily doped semiconductors \cite{efros1970},  Quantum Hall Effect \cite{raikh1993} and Weyl semimetals
\cite{rodionov2023} where the long-range correlations are attributed to Coulomb interaction.  An application of the Keldysh model to the theory of "liquid semiconductors" developed in the pioneering works of Sadovskii \cite{sadovskii1974,sadovskii1974a,sadovskii1979} paved an avenue for a description of the pseudogap phenomena well ahead the discovery of the high-temperature superconductors.  Physics of the pseudogap gave a new life to the multi-component Keldysh models
\cite{kuchinskii1999,sadovskii2001,schmalian1998,schmalian1999,chubukov2023a,
chubukov2023b,tchernyshyov1999}.  In these theories almost $\delta$-functional behaviour of the correlator has been controlled by the vicinity to the phase transition where Ornstein-Zernike theory is applicable.  The finite-size of the correlation length has been accounted by continuous fractions method \cite{sadovskii1974,sadovskii1979,schmalian1998,schmalian1999}. The number of component of the long-range model is equal to the spatial dimension ($d$$=$$2$) of the system. Besides, description of several polaron-related phenomena \cite{prokofev2026a,  prokofev2026b} greatly benefited from the methodology developed through the long history of studying the long-range disordered systems.

\begin{figure}
\centering
\includegraphics[width=70mm]{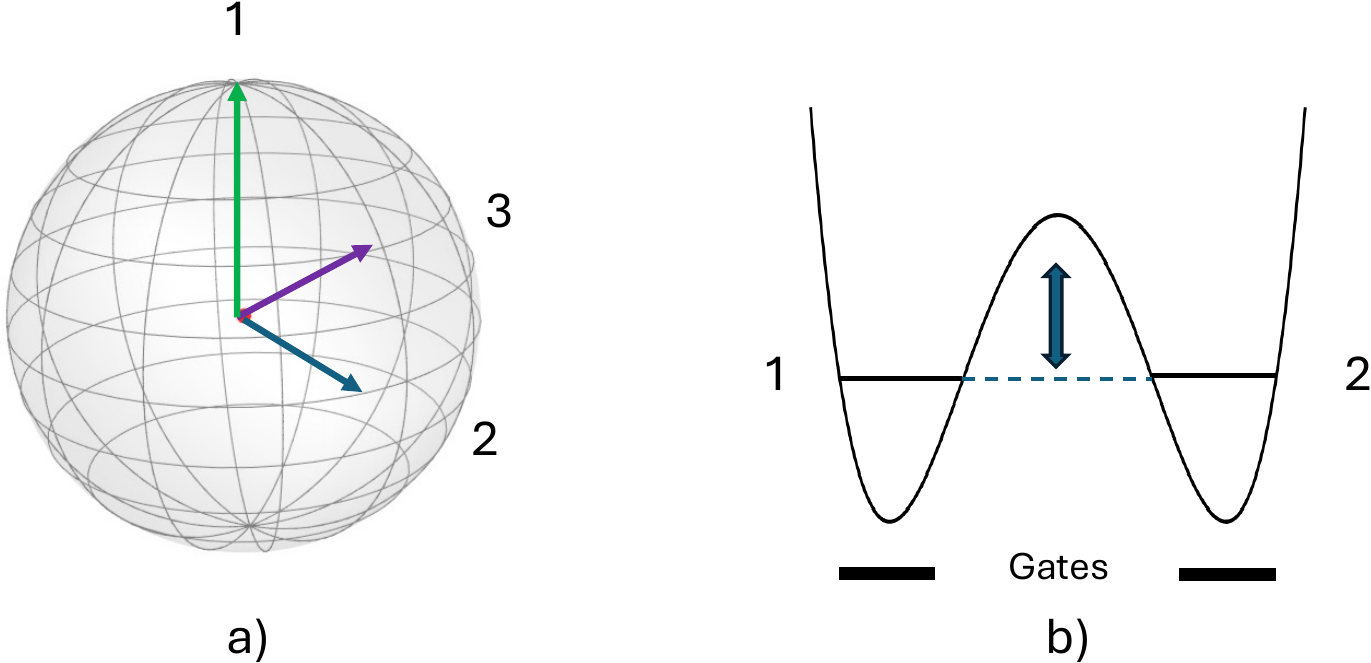}
\caption{Two realization of the multi-component random Gaussian field: a) spin dynamics in a random magnetic field: 1-Ising, 2-motion on an equator $S^1$, 3- motion on $S^2$; b) electron in a double-well potential with a noisy gates attached to both each well and fluctuating barrier. Solid lines inside the well illustrate the level position. Dashed line inside the barrier shows a tunnel path.}
\label{fig1}
\end{figure}

The idea of the Keldysh model construction {\it in the time domain} \cite{kiselev2009,kka2012,efremov2022,kuchinskii2024,sadovskii2026,
sadovskii2026a,sadovskii2026b}
opened a new direction for the dynamical Keldysh models analysis.  A dynamical random-field model describes behaviour of a dynamical system subject to the time-dependent Gaussian random potential  $v(t)$ defined by a color noise correlator:
\begin{equation}
D(t-t')=\langle v(t)v(t')\rangle =W^2 \exp\left(-\gamma|t-t'|\right).
\label{noise_corr}
\end{equation}
The coloured noise (\ref{noise_corr}) is defined by a characteristic time 
$\tau_{\rm noise}$$=$$1/\gamma$. There are two important limiting cases of (\ref{noise_corr}): i) $\tau_{\rm noise}$$ \to $$0$ or the white noise, describing Markovian dynamics  and ii) 
$\tau_{\rm noise}$$ \to $$\infty$  \cite{kiselev2009,  efremov2022, kuchinskii2024, sadovskii2026} totally antagonistic to the Markovian physics. The number of component of the dynamical Keldysh models does not depend any more on the number of spatial dimensions, but is controlled for example by the number of quantum dots in the setups where both quantum wells and barriers are manipulated by the noisy gates \cite{kiselev2009,  kka2012, efremov2022}. These models have been applied to the theory of Landau-Zener transition in noisy multi-level systems \cite{kiselev2013a,kiselev2013b} and central spin models \cite{kiselev2010,kiselev2012}.\\
\noindent
{\it {\color{blue}Time-evolution operator for the three-component Keldysh model.}}
In this {\it Letter} we consider a model defined by a Hamiltonian:
\begin{equation}
H_3=\Delta_1(t)\cdot \sigma^x+\Delta_2(t)\cdot \sigma^y+\Delta_3(t)\cdot\sigma^z.
\label{Ham3}
\end{equation}
Here $\sigma^\alpha$ are Pauli matrices and the classical random Gaussian field ${\bold\Delta}(t)$ is defined by its moments:
\begin{equation}
\langle \Delta_\alpha(t)\rangle\!=\!\Delta_\alpha^0,\;\langle [\Delta_\alpha(t) -\Delta^0_\alpha]
[\Delta_\beta(t')-\Delta^0_\beta]\rangle\! =\!\delta_{\alpha\beta}D(t-t')
\end{equation}
with a correlator $D(t-t')$ defined by (\ref{noise_corr}).  Since the model has unique energy scale $W$ we will use from now on the dimensionless variables, for example,
dimensionless time $t\to t\cdot W$, dimensionless energy $E\to E/W$ etc. Dimensionalization  is achieved by putting $W$$=$$1$ in the equations below. We use the system of units $\hbar$$=$$1$.  The $SU(2)$ model (\ref{Ham3}) describes the evolution of a spin $s$$=$$1/2$ in the three-component random Gaussian magnetic field or, equivalently, dynamics of a single electron in a double-well potential subject to the noisy gates acting on both the quantum wells and the inter-well barrier (see Fig.\ref{fig1}). 
The time evolution operator (TEO) satisfies an equation
\begin{equation}
i\frac{d}{d t}\hat{U}^\Delta=H_3\cdot \hat{U}^\Delta,\;\;\;\;\;\;
\hat{U}(0,0)=\hat {\mathbf{1}},
\end{equation}
which can be formally solved as
\begin{equation}
\hat{U}^\Delta(t,0)={\cal T}\exp\left(-i\int_0^t H_3(t') dt'\right).
\end{equation}
However, this exact formal solution expressed in terms of the time-ordered ${\cal T} $-exponent appears to be pretty useless for the Hamiltonians containing the non-commuting variables. A great simplification in construction of the TEO appears for the Hamiltonians possessing so-called {\it dynamical symmetries} (see the textbook \cite{dykhne1996}) where the Hamiltonians are given by a linear combination of a compact group generators with the time-dependent coefficients. Several representations for the simple exponential product form of the  TEO such as Magnus \cite{magnus1954}, Wei-Norman \cite{wei1963,wei1964}, Euler angles \cite{chalker2004,gritsev2013,ban1993} appear to be very useful. A standard approach to the random Gaussian field models with a white noise (zero correlation time) is based on either direct solution of the Langevin equations with multiplicative noise or derivation and solution  of  corresponding Fokker-Plank equation for the probability distribution. However, the strategy for the models with the long (infinite) time correlations is completely different: one has to find the TEO in {\it given realization} of the Gaussian disorder and average it over realization at the very last stage  \footnote{The approach is similar to the one used in the theory of glasses. In the case of a {\it quenched} disorder it is a logarithm of a partition function to be averaged over realizations. In the case of {\it annealed} disorder it is the partition function itself to be averaged.}.
We begin with the infinite time correlation models $\tau_{\rm noise}$$\to $$\infty$. By constructing the TEO and computing GFs we demonstrate afterwards how to account for a large but finite correlation time and moreover to crossover from the Keldysh model limit to the white-noise limit.
The Wei-Norman \cite{wei1963,wei1964} representation of TEO reads:
\begin{eqnarray}\label{wnk3}
\hat U^\Delta(t,0)\!=\!\exp\{f_{-}(t)S^-\}\!\exp\{f_z(t) S^z\}\!\exp\{f_{+}(t)S^+\}\;\;\;\;\;
\end{eqnarray}
where $S^{\pm}=S^x\pm iS^y$ and $\vec S =\frac{1}{2}\vec \sigma$ and three functions
$f_{\mp}$ and $f_z$ satisfy a system of coupled non-linear differential equations:
\begin{eqnarray}\label{wnk4}
\left\{
\begin{array}{lll}
\displaystyle i \partial_t f_{-}&=&\Delta^+-2\Delta_3 f_{-} -\Delta^- f_{-}^2\\
\displaystyle i \partial_t  f_z &=&2\Delta_3+2\cdot\Delta^- f_{-}\\
\displaystyle i \partial_t  f_{+}&=&\Delta^-e^{-f_z},
\end{array}\right.
\label{sysdif}
\end{eqnarray}
with initial condition $f_{-}(0)$$=$$f_z(0)$$=$$f_+(0)$$=$$0$ (see e.g.\cite{kka2012, dykhne1996}) and $\Delta^\pm$$ =$$\Delta_1$$\pm$$i$$\Delta_2$. The Wei-Norman representation for TEO  satisfies unitarity by its construction.

Parametrizing  
$\Delta^\pm$$=$$|\Delta|$$e^{\pm i\phi}$$\sin\theta$,
$\Delta_3$$=$$|\Delta |\cos\theta$
we obtain explicit solution of (\ref{sysdif}), see End Matter:
\begin{eqnarray}
f_{\mp}(t)&=& -ie^{\pm i\phi}\cdot\sin\theta\cdot\frac{\tan(|\Delta|t)}
{1-i\cos\theta\cdot\tan(|\Delta|t)},\nonumber\\
f_z(t)&=&\ln \left(\cos^2(\theta) \sin ^2(|\Delta|t)+\cos ^2(|\Delta|t)\right)-\nonumber\\
&-&2 i \arctan(\cos(\theta) \tan (|\Delta|t)),
\end{eqnarray}
where $|\Delta|$$=$$\sqrt{\Delta_1^2+\Delta_2^2+\Delta_3^2} $. Finally,
\begin{equation}
\hat U^\Delta(t,0)=
\begin{pmatrix}
e^{f_z/2}& f_{+}e^{f_z/2}\\
f_{-}e^{f_z/2}&\left[e^{-f_z/2}+f_{-}\;f_{+}e^{f_z/2}\right]
\end{pmatrix}.
\label{TEO3}
\end{equation}
This equation represents the TEO for $SU(2)$-invariant non-Abelian Keldysh model.\\
\noindent
{\it {\color{blue}  $\theta $$=$$\pi/2$:  Two-component Keldysh model.}}
For the two-component Keldysh model three functions defining Wei-Norman parametrization are greatly simplified:
\begin{eqnarray}
f_{\pm}(t)&=&-ie^{\mp i\phi}\tan(|\Delta_{12}|t),\nonumber\\
f_z(t)&=&\ln(\cos^2(|\Delta_{12}|t)),
\end{eqnarray}
and the TEO casts a much more simple form:
\begin{equation}
\hat U^\Delta(t,0)=
\begin{pmatrix}
\cos(|\Delta_{12}|t)& -ie^{-i\phi}\sin(|\Delta_{12}|t)\\
-ie^{i\phi}\sin(|\Delta_{12}|t)& \cos(|\Delta_{12}|t)
\end{pmatrix},
\label{TEO2}
\end{equation}
with $|\Delta_{12}|$$=$$\sqrt{\Delta_1^2+\Delta_2^2}$ 
\footnote{Factor $c$$=$$-i$ can be eliminated from the off-diagonal elements of the TEO by the re-definition of the polar angle for a zero-mean models only.}. 
This is the TEO of $U(1)$-invariant Abelian Keldysh model.\\
\noindent
{\it {\color{blue} $\theta$$ =$$0$ or $\theta$$=$$\pi$: One-component Keldysh model.}}
For the single-component Keldysh model yet another simplification occurs:
$f_-(t)$$=$$f_+(t)$$=$$0$, $f_z(t)$$=$$-2i|\Delta|t$ and the TEO is given by:
\begin{equation}
U^\Delta(t,0)=
\begin{pmatrix}
\exp\left(-i\Delta_3 t\right)& 0\\
0&\exp\left(i\Delta_3 t\right)
\end{pmatrix}.
\label{TEO1}
\end{equation}
In practice, only even part of the TEO $U^\Delta(t,0)$$=$$\cos(\Delta_3 t)\cdot \hat {\mathbf{1}}$ is contributing to the averaging over the Gaussian distribution of the one-component Abelian $Z_2$ disorder. \\
\noindent
{\it {\color{blue} Multi-component Keldysh models with a zero mean.}}
To begin with we  assume that $\Delta_\alpha^0$$=$$0$.  As the first non-trivial Keldysh model we consider the two-component $U(1)$ realization \footnote{The Gaussian distribution-averaged TEO for the single-component Keldysh model with zero mean is given by $\hat U_G(t) $$= $$\sqrt{2 \pi } e^{-\frac{t^2}{2}} \cdot \hat {\mathbf{1}}$. }.Referring to two illustrations shown on Fig.\ref{fig1} the models with the zero mean correspond for example to a random distribution of magnetic fields in a plane with the zero mean magnetic field (Fig.\ref{fig1}a) or a single electron dynamics in a double well potential (double quantum dot) with a barrier which is "closed" in average (hard wall) but "opened" due to the fluctuations.  For a symmetric double well potential the index $j$$=$$1,2$ labelling states in the first or second quantum well (see Fig. \ref{fig1}b) can be omitted. To obtain the averaged over disorder realization TEO we integrate  with Rayleigh distribution function:
\begin{equation}
P_R(|\Delta_{12}|)=|\Delta_{12}|\exp\left(-\frac{|\Delta_{12}|^2}{2}\right),
\label{rayleigh}
\end{equation}
assuming that the polar angle $\phi$ is uniformly distributed on a circle $S^1$ 
we write
\begin{equation}
\hat U_R(t,0)=\int_0^{2\pi}\frac{d\phi}{2\pi}\int_0^\infty d|\Delta_{12}| \hat U^\Delta(t,0)P_R(|\Delta_{12}|).
\end{equation}
Obviously,  the off-diagonal elements of the time evolution operator vanish after the integration over the polar angle.  Besides, the TEO is proportional to the unit matrix. Applying a Fourier transform to $\hat U_R (t,0)$ and defining the retarded single-particle GF as follows:
\begin{equation}
\hat G^R(t)=-i\theta(t)\langle \hat{U}^\Delta(t,0)\rangle_\Delta
\end{equation}
we finally (after averaging over the disorder realizations) obtain (the subscript in $G_2^R$
is attributed to $d$$=$$2$ component model and $\theta(t)$ is the Heaviside step function):
\begin{equation}
\hat G_2^R(\epsilon)=
\frac{e^{-\epsilon^2/2}}{2}
\left[
\epsilon\,\operatorname{Ei}\left(\frac{\epsilon^2}{2}\right)
-i\pi|\epsilon|
\right]\cdot \hat {\mathbf{1}}.
\label{kel_g2}
\end{equation}
Here $\operatorname{Ei}(u)$  is the exponential integral function 
\cite{abramovitz1965}.
Equation (\ref{kel_g2}) is identical to the two-component Keldysh model GF (see
\cite{kiselev2009, efremov2022}),  derived using Feynman diagrammatic technique
\footnote{The Green's function (\ref{kel_g2}) $G_2(\epsilon)$ satisfies a linear differential equation
$\frac{d G_2}{d\epsilon}
+\left(\epsilon -\frac{1}{\epsilon}\right)\cdot G_2=1$ \cite{kiselev2009, efremov2022}.}.
The imaginary part of the GF (\ref{kel_g2}) defines a density of states which is characterized by a  pronounced pseudogap dip evolving linearly at the small values of the energy.

To obtain the distribution-averaged TEO for the three-component $SU(2)$-invariant Keldysh model  with a zero mean we introduce the Maxwell-Boltzmann distribution function:
\begin{equation}
P_{MB}(|\Delta|)=\sqrt{\frac{2}{\pi}}|\Delta|^2\exp\left(-\frac{|\Delta|^2}{2}\right).
\end{equation}
Therefore, the distribution-averaged TEO is
\begin{equation}
\hat U_{MB}(t,0)\!\!=\!\!\int_0^{2\pi}\!\frac{d\phi}{2\pi}\!\int_0^\pi\! d\theta\sin\theta\!\int_0^\infty\! d|\Delta| \!\hat U^\Delta(t,0)\!P_{MB}(|\Delta|).
\end{equation}
The off-diagonal elements of the TEO vanish after the
integration over polar angle. similarly to the two-component case. 
The retarded GF is obtained  by
integrating the diagonal elements of (\ref{TEO3}) over the angle $\theta$ 
\footnote{The angle averaged TEO is 
$\langle \hat U_{MB}(t,0)\rangle_{\theta\phi}$$=$$\cos(|\Delta|t) \cdot \hat {\mathbf{1}}$.} following by the integration over the modulus of the fluctuating field and applying the Fourier transform:
\begin{equation}
\hat G_3^R(\epsilon)
=\left(\epsilon\left[
\epsilon\cdot {\cal F}_D\left(\frac{\epsilon}{\sqrt2}\right)-1
\right]
-i\sqrt{\frac{\pi}{2}}\,\epsilon^2e^{-\epsilon^2/2}\right) \cdot \hat {\mathbf{1}}.
\label{MB}
\end{equation}
Here ${\cal F}_D(u)$ is the Dawson function \cite{efremov2022, abramovitz1965}.
Equation (\ref{MB}) defines the $\theta$-$\phi$-angles-integrated retarded Green's function of the $d$$=$$3$ dynamical Keldysh model \cite{efremov2022}. The pseudogap in the density of states  is characterized by the $\epsilon^2$ at the small values of the energy.\\
{\it {\color{blue}Two-component Keldysh model with a non-zero mean.}}
We consider the Keldysh models with the non-zero mean. To illustrate how the methods developed in the {\it Letter} are used, we consider the two-component Keldysh model
\footnote{Calculations of the distribution-averaged TEO for the single-component Keldysh model with the non-zero mean are trivial and result in a shift of the Gaussian distribution peak.  Analysis of the  three-component model is more cumbersome but can be done along the lines of the method outlined in this {\it Letter}.}. The non-zero mean of the fluctuating Gaussian two-component field accounts for a small but finite transparency of the barrier for an electron in a double-well potential (Fig. \ref{fig1}b) or a non-zero mean value of the external magnetic field for an electron moving on a plane (Fig.\ref{fig1}a). We redefine the moments as follows \footnote{To simplify our notations we use specifically in this Section a definition ${\bf \Delta}_{12}\equiv \Delta$.}:
\begin{equation}
\Delta\!=\!\eta\! +\!\tilde\Delta,\;\langle\Delta\rangle\!=\!\eta, \;
\langle\Delta^\ast\rangle\!=\!\eta^\ast, \;
\langle\tilde\Delta(t)\tilde\Delta^\ast(t')\rangle\!=\!2D(t-t').
\label{finite1}
\end{equation}
Using the definition (\ref{finite1}) we obtain 
\begin{eqnarray}
&&|\Delta|^2 =|\eta|^2+|\tilde\Delta|^2+2|\eta\tilde\Delta|\cos(\alpha-\beta),\;
\eta =|\eta|e^{i\alpha},\\
&&\tilde\Delta=|\tilde\Delta|e^{i\beta},\;
\Delta =|\Delta|e^{i\phi},\;\tan\phi=
\frac{|\eta|\sin\alpha+|\tilde\Delta|\sin\beta}{|\eta|\cos\alpha+|\tilde\Delta|\cos\beta}.\nonumber
\end{eqnarray}
Without losing a generality we chose $\eta$ real and set $\alpha$$=$$0$
\footnote{The non-zero mean $\eta$ is a dimensionless variable normalized by $W$. One one hand, $\eta$ is related to the finite but exponentially small transparency of the barrier due to the under the barrier motion. On the other hand, the spectral function of the  equilibrium Johnston-Nyquist noise defines the temperature of the system $W^2$$\sim $$T$ (Boltzmann constant $k_B$$=$$1$). Therefore, in principle, 
dimensionless $\eta$ can be both $\eta$$ \ll $$1$ or $\eta$$ \gg $$1$  depending on the limiting cases of the model.}.
It corresponds to the case of the under-barrier tunnelling of electron in the absence of external magnetic field (Fig.\ref{fig1}b).  For the in-plane model on $S^1$ (Fig.\ref{fig1}a) it can be achieved by an appropriate re-definition (rotation) of the Cartesian axes in $x$$-$$y$ plane such as the non-zero mean magnetic field is directed along one of these axes.
Using equation for the two-component model TEO (\ref{TEO2})
and averaging TEO with the Rayleigh distribution for the two-component fluctuating field 
$\tilde{\bold \Delta}$  with a zero mean (\ref{rayleigh}) $P_R(|\tilde\Delta|)$ we get
\begin{equation}
\hat{\cal{U}}_R(t,0)=\int_0^{2\pi}\frac{d\beta}{2\pi}\int_0^\infty d|\tilde\Delta| 
\hat{U}^{\tilde\Delta, \eta}(t,0)P_R(|\tilde\Delta|).
\end{equation}
Here $\hat{U}^{\tilde\Delta, \eta}(t,0)$ is TEO (\ref{TEO2}) with a non-zero mean $\eta$
and zero-mean fluctuating field ${\bold{\tilde{\Delta}}}$.
The crucial difference of this case from the zero-mean models is the emergence of the off-diagonal correlations These  correlations describe the spin flip processes in the presence of a finite magnetic field (Fig.\ref{fig1}a) and account for finite probability  to tunnel in the absence of the fluctuations (Fig.\ref{fig1}b). Computing the integral over the polar angle for the diagonal matrix element of TEO we obtain
\begin{widetext}
\begin{equation}
u(\eta,t)=\int_0^{2\pi}\frac{d\beta}{2\pi}\int_0^\infty |\tilde\Delta|d|\tilde\Delta| 
\exp\left(-\frac{|\tilde\Delta|^2}{2}\right)
\cos(t\cdot\sqrt{\eta^2+|\tilde\Delta|^2+2\eta|\tilde\Delta|\cos(\beta)}).
\end{equation}
\end{widetext}
By computing function $u(\eta,t)$ and applying a Fourier transform \footnote{The calculations become simpler and more straightforward by proceeding through the following steps:
i) apply a Fourier transform to the time evolution operator; ii) integrate over polar angle $\beta$; iii) integrate  over  the zero-mean fluctuating variable $\tilde{\bold\Delta}$ using the Rayleigh distribution.}
we get
\begin{equation}
u(\eta,\epsilon)=\pi|\epsilon| e^{-(\eta^2+\epsilon^2)/2}I_0(\eta\cdot\epsilon)
\label{diag_u}
\end{equation}
Here $I_0(z)$ is a modified Bessel function of the first kind of the order zero \cite{abramovitz1965}. 

Due to the choice $\alpha$$=$$0$,  both off-diagonal matrix elements are identical and given by
\begin{widetext}
\begin{equation}
w(\eta,t)=-i\int_0^{2\pi}\frac{d\beta}{2\pi}\int_0^\infty |\tilde\Delta|d|\tilde\Delta| 
\sin(t\cdot\sqrt{\eta^2+|\tilde\Delta|^2+2|\eta|\tilde\Delta|\cos\beta})\cdot
\exp\left(i\arctan\left[\frac{|\tilde\Delta|\sin\beta}{\eta+|\tilde\Delta|\cos\beta}\right]-\frac{|\tilde\Delta|^2}{2}\right).
\end{equation}
\end{widetext}
Performing similar calculations for  $w(\eta, t)$ which defines 
the off-diagonal Green's function and applying a Fourier transform we obtain
\begin{equation}
w(\eta,\epsilon)=\pi\epsilon e^{-(\eta^2+\epsilon^2)/2}  I_1(\eta\cdot\epsilon),
\label{offdiag_w}
\end{equation}
where $I_1(z)$ is a modified Bessel function of the first kind of the order one \cite{abramovitz1965} \footnote{As both functions $I_0(u)$ and $I_1(u)$ have a large argument asymptote $I_n(u\gg1)$$\propto $$ e^u/\sqrt{u}$ \cite{gradstein2007,abramovitz1965}, the distribution starts to resemble centred at  $z$$=$$\pm\eta$ Gauss distribution for
$\eta z \gg 1$.  The energy splitting therefore is proportional to a tunnel matrix element.}.
With the help of these two functions (\ref{diag_u},\ref{offdiag_w}) describing statistics of Gaussian fields  (Rice distribution  \cite{rice1945}, see related work on localization in the random phase models \cite{sadovskii1981}) of the diagonal and off-diagonal matrix elements of the TEO we define two (diagonal and off-diagonal) GFs in the frequency domain. The diagonal retarded GF is given by
\begin{widetext}
\begin{eqnarray}
G^R_2(\epsilon,\eta) =\frac{1}{2}e^{-\eta^2/2}\int_0^\infty d z z e^{-z^2/2}I_0(\eta z)
\left[ \frac{1}{\epsilon-z+i 0^+}+\frac{1}{\epsilon+z+i 0^+}\right],
\label{greeng}
\end{eqnarray}
and the off-diagonal retarded Green's function casts the form
\begin{eqnarray}
F^R_2(\epsilon,\eta)= \frac{1}{2\epsilon}e^{-\eta^2/2}\int_0^\infty d z z^2 e^{-z^2/2}I_1(\eta z)
\left[ \frac{1}{\epsilon-z+i 0^+}+\frac{1}{\epsilon+z+i 0^+}\right].
\label{greenf}
\end{eqnarray}
\end{widetext}
These functions satisfy the system of coupled linear differential equations
\footnote{The system of coupled equations for the diagonal and off-diagonal Green's function resembles a system of Belyaev equations in the theory of superfluidity \cite{agd1963} or Gor'kov equations in the theory of superconductivity \cite{agd1963,sadovskii2019}. However, this similarity is misleading.  First, the equations remain differential both in the time and frequency domain. Second, the system of equations for the two-component Keldysh model is exact and does not assume any approximations such as for example a low density of a condensate or a mean-field theory. The system of equations can be derived by differentiating two exact Green's functions obtained from the time evolution operator or checked perturbatively using diagrammatics \cite{agd1963,sadovskii2019} by comparing the terms proportional to powers of a small parameter $\eta\ll 1$ order by order.}
\begin{eqnarray}
&&\frac{d G_2}{d\epsilon}
+\left(\epsilon -\frac{1}{\epsilon}\right)\cdot G_2
-\eta\cdot F_2=1,\nonumber\\
&&\frac{d F_2}{d \epsilon}+\epsilon\cdot  F_2 - \eta\cdot G_2 =0.
\label{diffu}
\end{eqnarray}
with a boundary condition at $\epsilon\to+\infty$:
\begin{equation}
G_2(\epsilon\to +\infty)=\frac{1}{\epsilon},\;\;\;F_2(\epsilon\to +\infty)=\frac{\eta}{\epsilon^2}.
\end{equation}
The equations (\ref{greeng},\ref{greenf}) and (\ref{diffu})  represent the central result of this {\it Letter}.\\
{\it {\color{blue} Multi-component Keldysh model with both non-zero mean and finite correlation time.}} 
To account for a large, but finite correlation time (finite non-zero $\gamma$ of the Gaussian disorder we use methods developed in \cite{sadovskii1974, sadovskii1979}. We demonstrate below how it works for the two-component Keldysh model with non-zero mean. Using Taylor expansion for the modified Bessel's functions \cite{gradstein2007, abramovitz1965} we re-express the bold diagonal and off-diagonal Green's functions  in terms of the infinite series with infinite convergence radius over even $2d$-component  Keldysh model Green's functions given by the continuous fractions:
\begin{eqnarray}
G^R_2(\epsilon,\eta,\gamma) &=&e^{-\eta^2/2}\sum_{k=0}^{\infty}\left[\left(\frac{\eta^{2}}{2}\right)^k \frac{G^R_{2k+2}(\epsilon,\gamma)}{k!}\right],\;\;\;\;\;\;\;\nonumber\\
F^R_2(\epsilon,\eta,\gamma) &=&\frac{\eta}{\epsilon}e^{-\eta^2/2}\sum_{k=0}^{\infty}\left[\left(\frac{\eta^{2}}{2}\right)^k\frac{G^R_{2k+4}(\epsilon,\gamma)}{k!}\right].\;\;\;\;\;\;\;
\end{eqnarray}
Here we defined 
\begin{equation}
 G_d^R(z,\gamma)
 =\cfrac{1}{z+i0^+-
 \cfrac{v_d(1)}{z+i\gamma-
 \cfrac{v_d(2)}{z+2i\gamma-
 \cfrac{v_d(3)}{z+3i\gamma+-\ddots}}}}.
\end{equation}
the Green function of the zero-mean finite-correlation time spherical $d$-component models on the $S^{d-1}$ with a surface $S^{d-1}=\frac{2\pi^{d/2}}{\Gamma(d/2)}$ and  volume $\Omega_d=2^{d/2-1}\Gamma\!\left(\frac d2\right)$,  $\Gamma(z)$ is Euler's Gamma-function
\footnote{The infinite noise correlation time Green's function of a $d$-component spherical Keldysh model is given by $ G_d^R(x)
 =\frac{1}{2\Omega_d}\int_0^\infty du\,u^{d-1}e^{-u^2/2}
 \left[
 \frac{1}{x-u+i0^+}+\frac{1}{x+u+i0^+}
 \right]$.}.
For the even numbers of components, $d$$=$$2m$, combinatoric coefficients $v_{2m}(k)$ are given by simple expressions \cite{fatima2026,schmalian1998,schmalian1999}:
\begin{equation}
 v_{2m}(2n-1)=2(n+m-1),
 \qquad
 v_{2m}(2n)=2n.
\end{equation}
Combining all parts of the solution we finally get the Green's function  of the two component Keldysh model with both finite correlation time and the non-zero mean.
\begin{figure}[t]
\centering
\includegraphics[width=65mm]{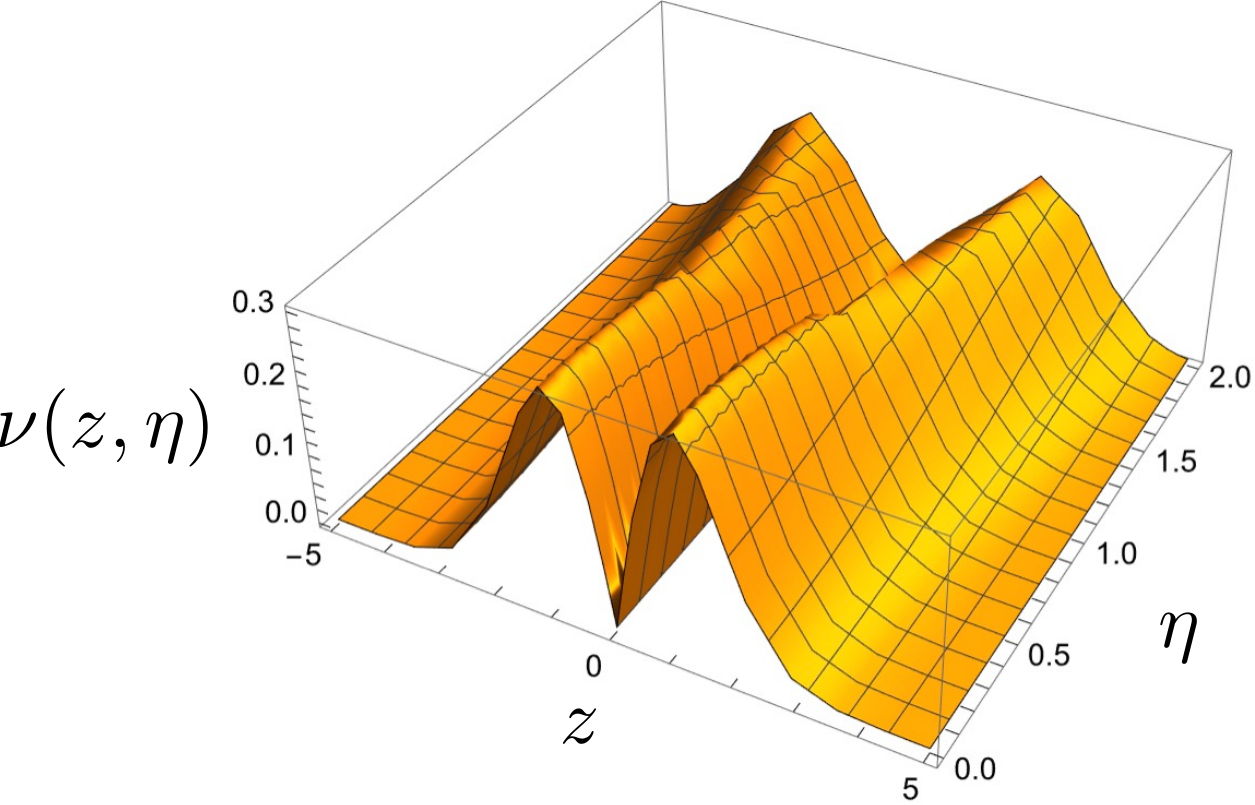}\\
\includegraphics[width=65mm]{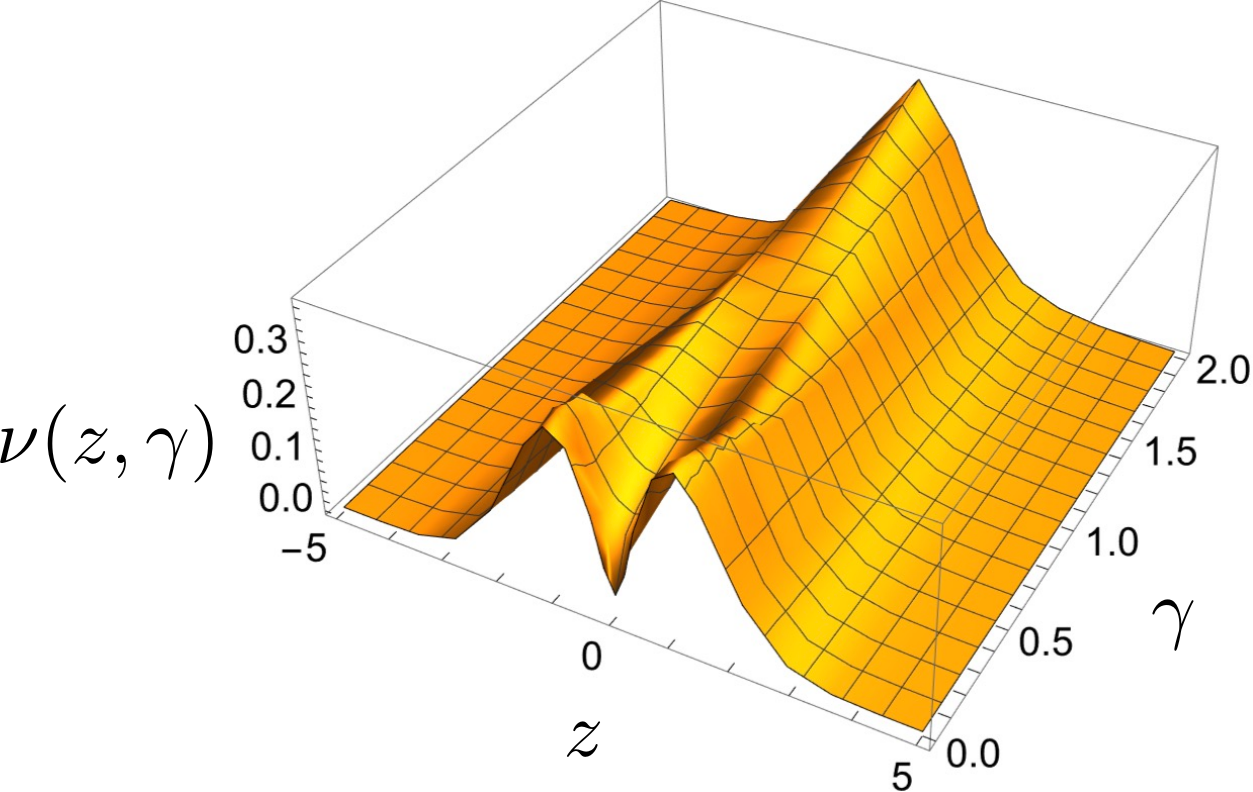}
\caption{Density of States (DoS) 
$\nu(z)$$=$$-\frac{1}{\pi}$${\rm Im}$$G_2^R(z)$ of the two-component Keldysh model as a function of dimensionless energy $z$$=$$\epsilon/W$.  Top panel: effects of the finite mean transparency of the barrier $\eta$.  Bottom panel: influence of the finite relaxation rate 
$\gamma$ (inverse $\tau_{\rm noise}$).}
\label{fig2}
\end{figure}
The impact of these two variables on a density of states (DoS) with a pseudogap is different \cite{kiselev2027a,
kiselev2027b}: finite $\gamma$ results in a non-zero DoS at zero energy (incomplete orthogonality) reducing the pseudogap "from the bottom" (Fig.\ref{fig2} bottom)
\footnote{Finite rate $\gamma$=$1/\tau_{\rm noise}$ brings a new dimensionless parameter $W\cdot \tau_{\rm noise}$ to the model: Keldysh limit corresponds to $W\cdot \tau_{\rm noise} $$\gg $$ 1$ and defines Gaussian shape of the DoS peaks and tales. In the opposite limit $W\cdot\tau_{\rm noise}$$  \ll $$1$ the line shape is Lorentzian.}. 
The finite mean value $\eta$ leads to the reduction of the peak's hight (Fig.\ref{fig2} top)
due to the re-distribution of the spectral weight between the diagonal and the off-diagonal Green's functions modifying the pseudogap "from the top".  As a result, both effects lead to the reduction of the pseudogap.  At some critical value of the $\gamma$ the pseudogap vanishes changing behaviour of DoS from the central dip to the central peak. The line $\gamma(\eta)$ defines the boundary between two phases: two peaks separated by the dip vs one central peak (see Fig.\ref{fig2} and discussion in \cite{kiselev2027b}). \\
\noindent
{\it {\color{blue} Discussions and prospectives.}} The Wei-Norman parametrization applied for the construction of the time-evolution operator for the dynamical non-Abelian $SU(2)$ model of the random Gaussian disorder (multi-component Keldysh models) paves an avenue for finding exactly the time evolution operator for dynamical models  beyond $SU(2)$, symmetry for example $SU(3)$ 
\cite{dattoli1987,kka2012, kiselev2013a, kiselev2013b,kiselev2026a} and $SU(1,1)$ \cite{dattoli1987}.  Moreover, it provides a very power tool for the analyse of the models with dynamical symmetries on various manifolds including but not limited by $CP^N$
\cite{kiselev2026a}. Another unsolved problem is related to construction of a two-particle Green's function \cite{sadovskii1991} for the multi-component models and, in particular,
solution of a two-particle singlet-triplet case described by $SO(4)$ symmetry.
Special interest deserve the non-Abelian models with the non-zero mean where the off-diagonal ("anomalous") Green's functions emerge. It is important, in particular, in connection with the superconductivity, spin and charge density waves and excitonic instability  \cite{kuchinskii2024,sadovskii2026,sadovskii2026a,sadovskii2026}, the off-diagonal correlations in a system of coupled quantum dots \cite{kka2012, efremov2022}. including Landau-Zener transitions \cite{kka2012,kiselev2013a,kiselev2013b} and the spectral collapse \cite{gefen2002}. Analysis of the dynamical models with the Gaussian disorder characterized by the large but finite noise correlation time poses also new questions, for example, related to the Ward identities \cite{kuchinskii2024,sadovskii2026} and continuity equations.\\
\noindent
{\it {\color{blue} Conclusions.}} Summarizing, we constructed exact solution for dynamical model with a classical random Gaussian disorder on a $d$$-$$1$-dimensional sphere $S^{d-1}$. The solution includes a non-Abelian $SU(2)$ symmetry class $d$$=$$3$ and two Abelian classes $U(1)$ for $d$$=$$2$ and $Z_2$ for $d$$=$$1$. While the Green's functions of the models with zero mean of the fluctuating field obtained using exact form of the time evolution operator coincide with the known results derived by means of the Feynman diagrammatics by either applying the Borel 
re-summation \cite{keldysh1965, kiselev2009} or by solving Dyson equation \cite{kiselev2009}, the Green's functions for the multi-component $d$$>$$1$  models with the non-zero mean represents a new result. We derived the exact form of the diagonal and the off-diagonal Green's functions and obtain the system of linear differential equations which these GF satisfy.  We illustrate the solution using a paradigmatic $d$$=$$2$ dynamical Keldysh model. The results reported in this {\it Letter} can be used for a wide range of the dynamical models: from the quantum dynamics of coupled quantum dot prototyping various qubit's realizations and numerous setups experimentally accessible with ultra-cold gases to the out-of-equilibrium superconducting and magnetic instabilities.
\\
\noindent
{\it {\color{blue} Acknowledgements}} M.N.K. is grateful to Beno\^{i}t Dou\c{c}ot and Yuval Gefen for illuminating discussions and M.V. Sadovskii for useful comments and interest to this work. We appreciate collaboration with Fatima Zohra Bendjeddou and Dmitriy Efremov on closely related projects. M.N.K. is also grateful for the hospitality to IFW Dresden where this project was completed. This research was supported in part by grant NSF PHY-2309135 to the Kavli Institute for Theoretical Physics (KITP). 
\noindent
{\it {\color{blue} Data availability}} No data were created or analysed in this study.
\bibliography{DynKeldTEO}
\appendix
\section*{End Matter}
\setcounter{equation}{0}
\renewcommand{\theequation}{A\arabic{equation}}
We provide some details for derivation of functions defining the time evolution operator.
The first equation is a non-linear Riccati equation
\begin{equation}
i \frac{d f_{-}}{d t}=\Delta^+-2\Delta_3 f_{-} -\Delta^- f_{-}^2,\;\;\;\;f_-(0)=0
\end{equation}
which can be transferred to the ordinary second order linear differential equation
\begin{equation}
\frac{d^2 u}{d t^2}-2i\Delta_3\frac{d u}{d t}+\Delta^+\Delta^- u=0
\end{equation}
by a substitute
\begin{equation}
f_{-}(t)=\frac{i}{\Delta^-}\frac{d\ln u}{d t}=\frac{i}{\Delta^-}\frac{1}{u}\frac{d u}{d t}
\end{equation} 
Applying another substitute $u(t)=v(t)\cdot e^{i\Delta_3 t}$
we get
\begin{equation}
\frac{d^2 v}{d t^2}+|\Delta|^2\cdot v=0
\end{equation}
where we denoted
$|\Delta|^2=\Delta_1^2+\Delta_2^2+\Delta_3^2$. This equation describes a clockwise and counter clockwise spin precession with a Larmor frequency $\Omega_L$$=$$|\Delta|$ on $S^2$  in a given realization of the random Gaussian magnetic field  ${\bf \Delta}$.

To find $f_z(t)$ we solve the differential equation
\begin{equation}
\frac{d f_z}{d t}=-2i\Delta_3-2i\Delta^{-}\cdot f_{-}(t),\;\;\;\;\; f_z(0)=0.
\end{equation}
The solution subject to initial condition is given by the integral
\begin{equation}
f_z(t)=-2i\Delta_3\cdot t -\frac{2\cdot \Delta^+\cdot \Delta^-}{|\Delta|^2}
\int_0^{|\Delta|t} d x\frac{\tan x}{1-\displaystyle\frac{i\Delta_3}{|\Delta|}\tan x}
\end{equation}
The third non-linear differential equation  for $f_{+}(t)$ reads
\begin{equation}
\frac{d f_+}{d t}=-i\Delta^-e^{-f_z},\;\;\;\;\;\;\;\;\;f_+(0)=0
\end{equation}
with a solution
\begin{equation}
f_{+}(t)=-ie^{-i\phi}\sin(\theta)
\int_0^{|\Delta| t} d x
\frac{\exp\left[2 i \arctan(\cos(\theta) \tan (x))\right]}
{\cos^2(\theta) \sin ^2(x)+\cos ^2(x)}
\end{equation}
Here we used an identity
$$
\arctan(\cos(\theta) \tan (|\Delta|t))=
\frac{1}{2i}\ln\left[\frac{\cos(|\Delta|t)+i\cos\theta\sin(|\Delta|t)}{\cos(|\Delta|t)-i\cos\theta\sin(|\Delta|t)}\right]
$$
\end{document}